# Forty Years of Linking Variable Star Research with Education

John R. Percy[1]

[1]Department of Astronomy & Astrophysics, and
Dunlap Institute for Astronomy & Astrophysics, and
Centre for STEM Education
University of Toronto

**Abstract:** In this review, I reflect on four decades of my experience in linking astronomy research and education by supervising variable-star research projects by undergraduates, and by outstanding senior high school students. I describe the evolution of my experience, the students I have supervised, the nature of their projects, the educational contexts of the projects, the need for "best practices", the journals in which we publish, and the special role of the American Association of Variable Star Observers (AAVSO). I then describe our recent research on pulsating red giants and related objects, including three astrophysical mysteries that we have uncovered. Finally, I suggest how my projects might be scaled up or extended by others who supervise student research.

## Introduction

Variable stars are stars which change in brightness, which they may do for many reasons. I began my research career as a theoretician, modelling stellar structure, evolution, and pulsation of stars. Very soon, I also developed an interest in the observation of variable stars, thanks to mentors such as Don Fernie. I already had a strong interest in education, so it was not surprising that I soon began supervising undergraduate student research projects, before many of my colleagues realized that undergraduate students can participate meaningfully in research.

Undergraduate research has been a priority for North American universities and colleges for many years, and that has certainly been true for my university. Nowadays, an even higher priority is given to student "engagement", and research is one way to engage students. It gives them real-world skills and experience. It may help them to decide whether or not they would like a career in research. They can interact personally, one-on-one with a researcher/professor. Not least: they have someone to write letters of reference for them!

Initially, my students did photoelectric photometry (PEP) of variable stars in the summer, using an on-campus 40 cm telescope -- despite the challenges of doing so in the centre of a climatically-underprivileged city like Toronto. Forty years ago, in 1977, I joined the American Association of Variable Star Observers (AAVSO). The AAVSO is a non-profit society which, since 1911, has encouraged, coordinated, evaluated, compiled, processed, archived, published, and freely disseminated variable star observations around the world. Its website is a "gold mine" of public data, resources, and information, for students and other researchers. It includes over 33 million measurements, made mostly by skilled amateur astronomers – "citizen scientists".

AAVSO data share some characteristics with data from robotic telescopes. The observations are guided by well-established AAVSO observing programs, and those of its observing "sections", and by "campaigns" on specific stars, using standard procedures. The user may forget that they come from humans, not from robots. But each observation comes with an "observer code" attached to it, belonging to a real human observer.

Through the AAVSO, I met Director Dr. Janet Mattei. We had much in common, including a strong interest in education. In the early 1980's, we got a grant from the Toronto-based Bickell Foundation, to do a series of projects with AAVSO visual observations of 391 Mira stars e.g. Percy & Colivas (1999). At the same time, PEP was on the rise, so Janet and I created the AAVSO PEP program (Percy, 2012). AAVSO visual and PEP data were ideal for my students, and still are.

Janet and I independently realized that variable star astronomy had even greater educational potential. In the 1990's, with support from the National Science Foundation, we created *Hands-On Astrophysics (HOA;* Mattei & Percy, 2000). With the help of our colleague Donna Young, *HOA* morphed into *Variable Star Astronomy:* www.aavso.org/vsa. Education remains one of the AAVSO's highest priorities; for some resources, see www.aavso.org/education. We also have a list of suggestions for student projects: www.aavso.org/student-observation-projects.

Robotic telescopes were also on the rise, and I obtained PEP data from the Fairborn Observatory telescopes, and later collaborated with robotic-telescope user Greg Henry, Tennesee State University, e.g. Percy et al. (2001). My students also made good use of the extensive CCD data on T Tauri stars amassed by Bill Herbst and his (human) students at Wesleyan University. We also used photometry from the *Hipparcos* satellite (Percy et al., 2004). Why? Because the data were there, and contained useful science. Nowadays, almost all of our projects are done with AAVSO archival visual and PEP data. The visual data are unique in that they go back a century or more, and provide the only way of studying the behavior of variable stars on a very long timescale.

**Our Projects**

My strategy is to give each student (or two, or three) a small, self-contained, original research project which can be completed in the time available, and published in a refereed journal with the student(s) as co-author. The students develop and integrate their science, math, and computing skills, motivated and engaged by doing and publishing real science, with real data. One important consideration is how to guide the research, while allowing the students some initiative, creativity, and flexibility.

Students spend the first few weeks reading, meeting with me, and discussing topics that they want to understand better. The discussions often go far beyond astronomy! They start with my book (Percy, 2007), and information on the AAVSO website, and work up to papers in research journals such as the *Journal of the AAVSO (JAAVSO)*. In this way, the students learn to plan and strategize, and build up confidence that they can actually do research. They are then introduced to the data and software that they will use, run the software on sample data, and discuss the results with me. They soon encounter the concepts of measurement and measurement error, graphs such as light curves, periodicity and irregularity, and time-series analysis in general. Throughout, they learn about record-keeping, and how to communicate with another scientist – me. In the end, they may write a project report. If so, I read one or two drafts, to give them practical guidance, experience, and feedback on how to write a scientific paper.

Over the years, we have carried out a wide variety of projects, on a wide variety of variable star types. Our publications are listed in the SAO/NASA Astrophysics Data System (ADS). For a few years, we worked on Be stars, including as part of multi-longitude, multi-wavelength campaigns (Percy et al., 1996, 2004). We worked on T Tauri stars (Percy et al., 2010), sometimes in collaboration with Bill Herbst. But for the last two decades, we have concentrated on pulsating red giants (PRGs): PEP searches for and studies of small-amplitude PRGs, classification of hundreds of PRGs in the AAVSO archive (e.g. Percy & Terziev, 2011; most were non-variable), studies of multiperiodic PRGs, and studies of pulsating red supergiants (PRSGs), whose variability is on such a long time scale that it can only be studied using the visual observations in the AAVSO archive.

Our current projects, described below, use AAVSO archival visual and PEP data on PRGs and PRSGs, and the AAVSO VSTAR software package for Fourier and wavelet analysis. As the students begin to use the data and software, they soon learn that time-series analysis is a "black art". The stars' variability occurs on several timescales, and is not always periodic and sinusoidal. The data are not always dense, and have seasonal and sometimes monthly gaps which produce artifacts in the Fourier spectrum (Percy, 2015). Understanding the complexities of the stars, and the limitations of the data, is at the core of the project. Once the students have understood the results on each star, they have to understand what the whole ensemble of program star results is telling them. This requires scientific judgement, judicious use of graphs and tables, and the ability to understand and express these results in words.

**Where We Publish**

The results of any original, significant research project should be published, ideally in a widely-available refereed journal. Refereeing provides an additional level of "quality control". Over the years, we have published mainly in two journals: the *Publications of the Astronomical Society of the Pacific (PASP)*, and *JAAVSO*. Stars are one of the specialties of the *PASP* and, as a former president of the ASP, I like to support this very worthy organization. But *PASP* has page charges and, since I gave up my research grant ten years ago when I "retired", I now publish almost exclusively in *JAAVSO*, which does not have page charges for AAVSO members.

*JAAVSO* is an on-line refereed journal which, after a short proprietary period, is publicly available. It publishes papers on variable stars and related topics, including their use in education. The *JAAVSO* website includes a wealth of information about how to write, format, and publish scientific papers, as well as lots of examples. Currently, I am the editor. For papers submitted by me or my students, the refereeing process is handled by our Production Editor, Dr. Michael Saladyga. The refereeing is "blind"; the referee does not know who the authors are – at least in theory!

One might question whether students should be co-authors of research papers, or whether they should simply be acknowledged. In principle, all authors should understand the content and implications of a paper, as well as contributing significantly to it. I suspect that my students could do this, especially if they had a few days to study and review!

**Educational Contexts**

For me, these have ranged from the standard fourth-year (senior) thesis, to various opportunities for undergraduate research, to a unique program for outstanding senior high school students, whose capabilities were very similar to those of undergraduates. See Percy (2008) for an overview of my student research experiences. Presently, I am supervising undergraduates through a work-study program. For many years, it was funded by the provincial government. Now, it is funded by my university, though faculty are expected to cover a small fraction of the cost. It's a win-win-win situation: students get paid; they get career-related experience; and good science gets done and published.  Participation in work-study is officially recorded as a co-curricular activity on the students' formal academic transcript.

For many years, my department's summer undergraduate research program (SURP) was standard one-on-one supervised research.  Recently, it has been enhanced to include mini-lectures, team-building workshops, opportunities to read and interpret research papers, training and experience in abstract-writing, report-writing and presentation, and training and opportunities for public outreach – which is a high priority for us.  Because the University of Toronto astronomical community includes institutes devoted to theory, and to instrumentation, as well as to general astronomy, SURP students are exposed to almost every branch of astronomy and astrophysics.  SURP students also have access to the lectures in our annual Summer School on Astronomical Instrumentation for senior undergraduates and beginning graduate students, apparently the only one of its kind.  See Amaral & Percy (2016) for an excellent description of SURP from the point of view of a student.

Ariel Amaral was a summer work-study student who had developed an interest in red giant stars as a result of her third-year astrophysics course (she got an A+ on the course).  She decided that, rather than doing a small, focussed project, she wanted to sample various ways of using variable stars to understand these stars.  At the same time, *JAAVSO* was interested in publishing more papers on education, so she decided to do the project and write the paper as a combined education and research project, from the student's perspective.  I recommend it highly.

The Research Opportunity Program is a second-year (sophomore) program at my university, in which students can participate in a faculty member's research for full-course credit.  Because a grade is assigned, I structure the assessment with credit given for a mid-term report, a  final report,  a presentation, a final oral interview, and for the student's general engagement in the project.  I give the students lots of advance guidance and feedback on how they will be assessed.  The highlight of the program is the year-end "research fair", at which students from across the Faculty of Arts and Science present their projects.

The University of Toronto Mentorship Program was a prestigious, highly-competitive program which, for 25 years,  enabled outstanding senior high school students to work on research projects at my university (Percy, 2006).   In almost every case, it was voluntary for the students though, in a few cases, they got a co-op credit, or submitted their work to their school as an Independent Studies course.  I found that there were advantages to accepting more than one student; two to four students could work effectively as a team. Because of my long-time interest and involvement in school liaison, the program was also a useful connection with the schools.

Most students analyzed archival variable-star data as described above, but I always encourage students to connect their research project to their other interests.  One student took a time-series of 35mm slides (remember those?)  of two bright variable stars from her backyard, and used these for an in-school science class activity.  Several students presented their project to their class, or to my astronomy department, or even to a conference.  One student attended a meeting of the AAVSO in Cambridge MA, and wrote up her experience in the *AAVSO Newsletter*.  A few students wrote useful software, or created webs resources for student research.  They even developed a "new" method of time-series analysis – self-correlation.  At the end of the year, there was a "research fair" at which the students presented the results of their projects (Figure 1).

Sadly, the mentorship program was discontinued after 25 years, because not enough of the students were enrolling at the University of Toronto! It was not a very effective recruitment tool, but it was certainly beneficial for the students and the supervisor

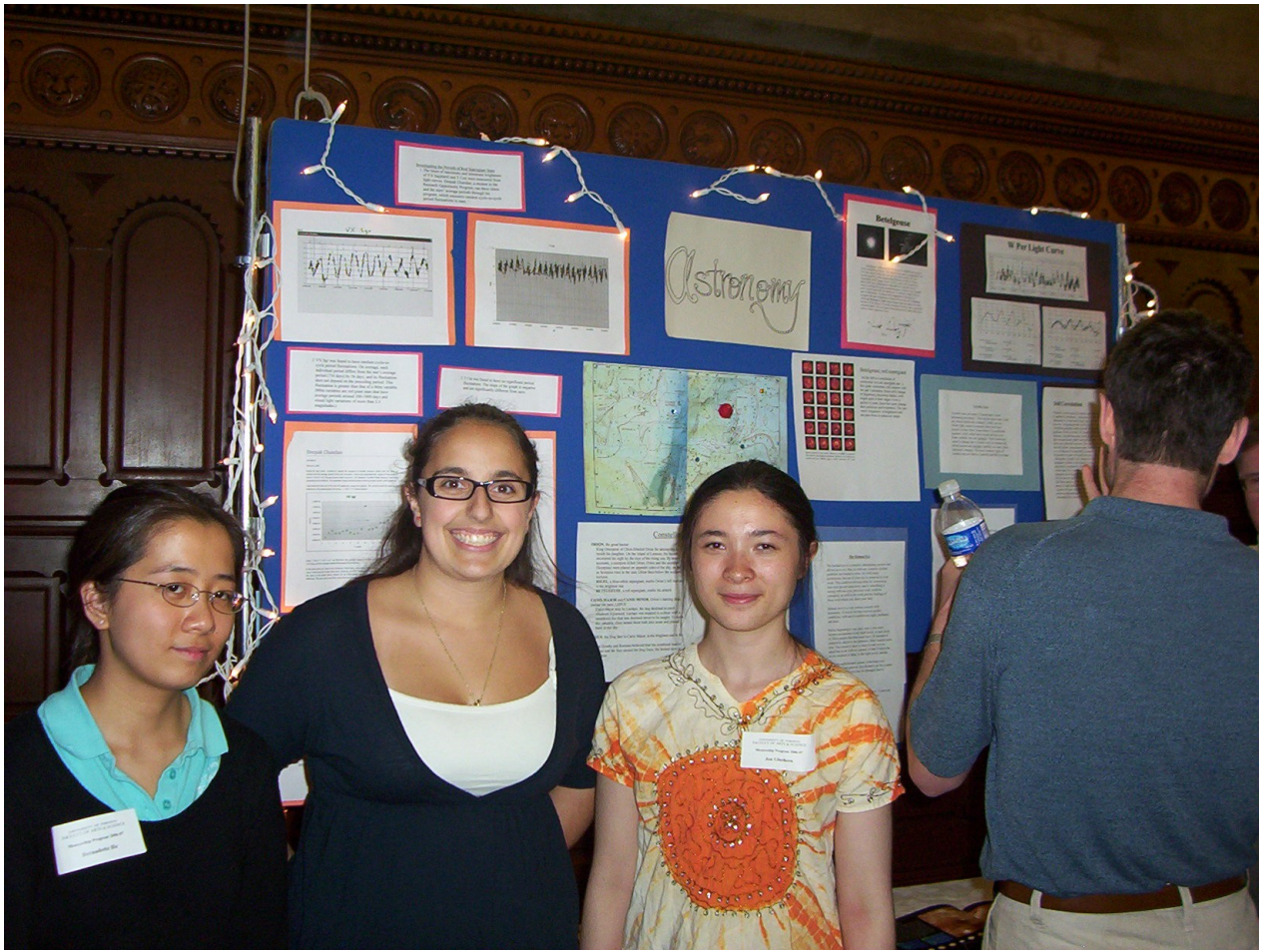

Figure 1. Bernadette Ho, Elena Favaro, and Jou Glasheen with their poster at the 2006-2007 University of Toronto Mentorship Program Research Fair, attended by teachers, family and friends, mentors and mentees. The white Christmas lights add a festive and somewhat astronomical touch.

**Educational Considerations**

As an educator, I must stress the need for "best practices" in every educational activity. If you create a research program for students, at any level, be sure to build in a process for rigorous evaluation and constant improvement. Be sure that the program is consistent with curriculum objectives, and provides students with a full range of professional development "hard" skills and "soft" skills. And be sure that supervisors are well-trained – not just in their research, but in their teaching, supervision, and mentorship. Remember that, whereas schoolteachers receive at least a year of training, university instructors usually receive none!

**About My Students**

In the last 20 years (I was too lazy to count back further), I have supervised 65 undergraduates and 44 high school students. They are about evenly split between male and female; about half are visible minorities.  They have included one Rhodes Scholar (Wojciech Gryc), a Space-X rocket scientist (Margarita Marinova), an eminent Harvard geoengineer (David Keith: "Dr. Cool"), one of *Time* magazine's "Hundred Most Influential People" (climate scientist Kathy Hayhoe), a Gruber Cosmology Prize winner (Wendy Freedman), a university dean (Doug Welch), a few other professional astronomers, some doctors, engineers, computer scientists, educators – even a golf pro.  I am very proud of them all.  My wife (Professor Emerita of Physiology) and our daughter (Professor of English) also supervise students in the same programs as I do, with great success.  One of the highlights of my academic career was giving a joint presentation with them, at our university's annual Teaching and Learning Symposium, on "Engaging Students in Research and Scholarship: Three Perspectives".

**Projects with Red Giants**

When low-mass stars (typically 0.6 to 10 times the mass of the sun) start to run out of hydrogen thermonuclear fuel, they swell up and become *red giants* – becoming larger, brighter, and cooler.  They then ignite helium thermonuclear fuel in their core.  As they run out of this fuel, they again become larger, brighter, and cooler; they are *asymptotic-giant-branch* (AGB) stars.  Both red giant and AGB stars are thermodynamically unstable to radial pulsation.  Initially, their periods are a few days and their amplitudes are a few hundredths of a magnitude.  At their most extreme sizes, their periods are hundreds of days, and their visual amplitudes are several magnitudes.

Since the pulsation period of a star correlates with its luminosity, PRGs can be used as "standard candles".  Slow changes in their pulsation periods can be used to detect and measure their evolution (Neilson et al., 2016).  Some PRGs pulsate in two or more modes; these provide information about the star's mass and internal structure.  Large-amplitude pulsation drives mass loss in extreme PRGs; this is what ultimately terminates the star's AGB evolution.  And unexplained phenomena in their pulsation reflect poorly-understood or unknown processes, as the following three examples illustrate.

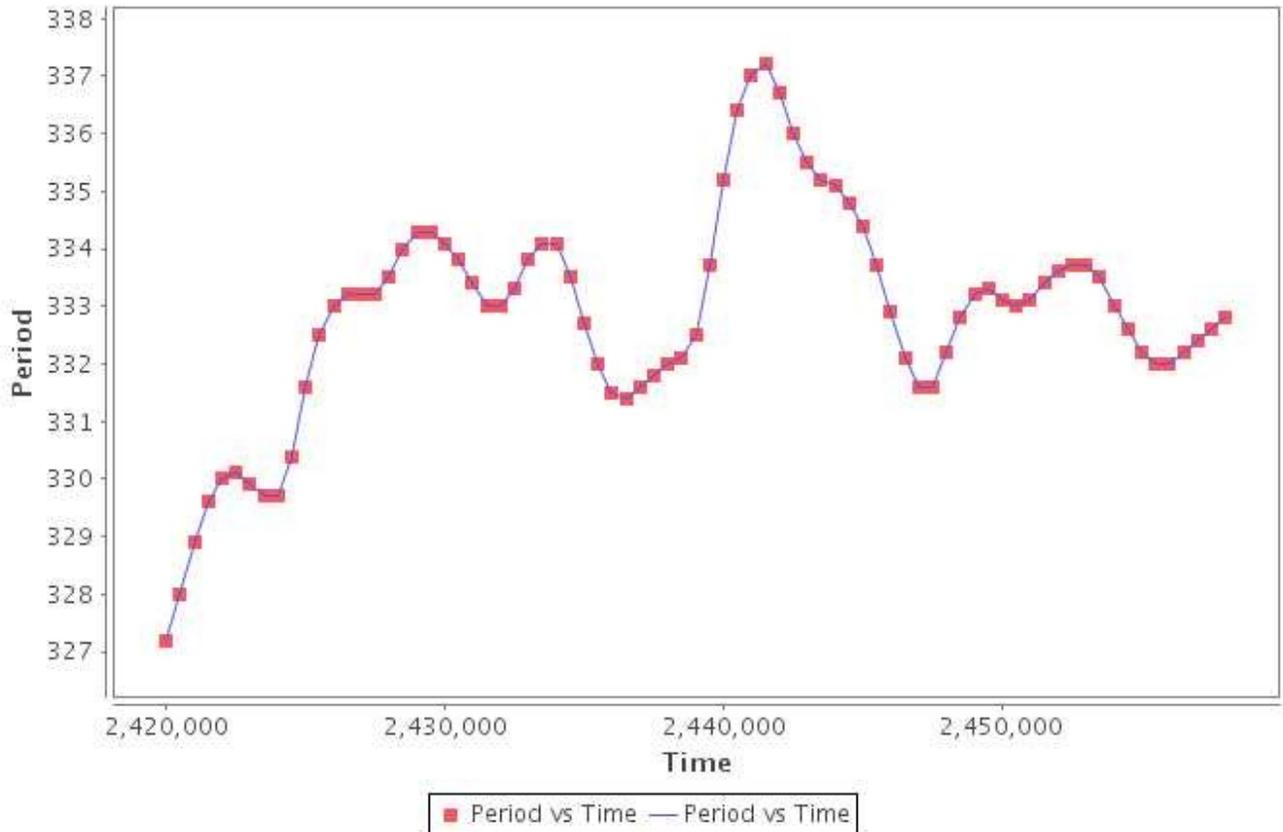

Figure 2: The "wandering" period in days of Mira, the prototype of large-amplitude PRGs, as a function of Julian Date. All Miras show this behavior (Percy & Colivas, 1999), which can be modelled by random cycle-to-cycle fluctuations (Amaral & Percy, 2016), perhaps due to the effects of giant convective cells.

*Random Cycle-to-Cycle Period Fluctuations*

The pulsation periods of PRGs, as determined by wavelet or (O-C) analysis, do not remain constant. A small number change systematically and significantly, due to "helium shell flashes" in the stars. The periods of all other PRGs wander with time in a "random walk" fashion (Figure 2). Almost a century ago, Eddington & Plakidis (1929) showed that, for a few such stars, this could be modelled as random, cycle-to-cycle period fluctuations. Percy & Colivas (1999) showed that this held true for 391 Mira stars in the AAVSO visual observing program; see also Amaral & Percy (2016). Although we assume that the fluctuations are caused by giant convection cells, we have no proof for this.

*Long Secondary Periods (LSPs) in Pulsating Red Giants*

Systematic photographic and visual observations showed, a century ago, that about a third of PRGs have a secondary period, 5-10 times longer than the pulsation period. This phenomenon was re-discovered, and promoted by Peter Wood (2000), based on large-scale survey observations of the Large Magellanic Cloud as part of the MACHO project. Despite extensive research by Wood and others, the cause of LSPs remains unknown.

My students and I have been mining the AAVSO's century-long database of visual observations of PRGs and PRSGs, along with PEP observations of shorter-period stars e.g. Percy & Deibert (2016). Figure 3 shows an example.   We want to use these databases to establish everything that we can about

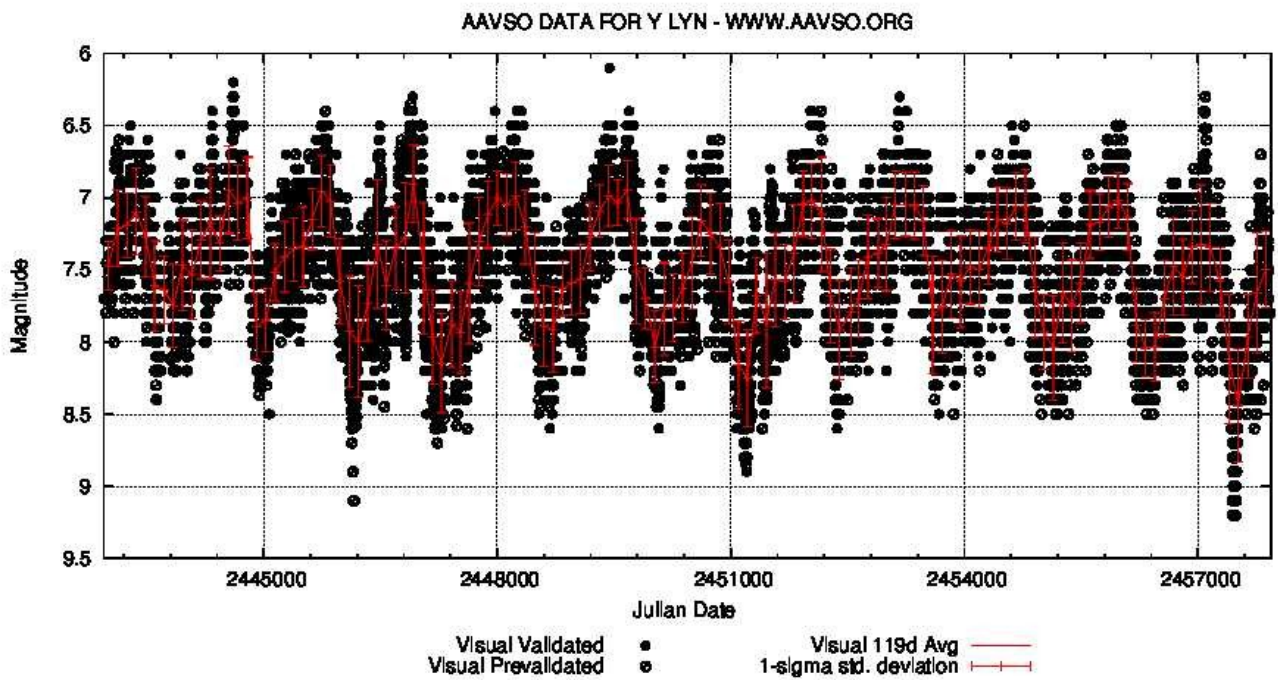

t

Figure 3:  The light curve of Y Lyn, showing the long secondary period (red).  AAVSO visual observations have been averaged in bins of 119 days (the pulsation period) to emphasize the LSP (Percy & Deibert, 2016).  The LSP is 1259 days.

this phenomenon, in the hope that it will help to solve what Wood considers to be the most important unsolved problem in stellar pulsation.

*Amplitude Variations in Pulsating Red Giants*

Unlike the two phenomena described above, the discovery that virtually all PRGs show significant changes in amplitude – up to a factor of 10 -- seems to be a recent discovery by Percy and Abachi (2013).  The timescales of these amplitude variations are about 20-30 pulsation periods.  The same is true for PRSGs (Percy & Khatu, 2014) and yellow supergiants (Percy & Kim, 2014).  We continue to study this phenomenon.  Its cause remains unknown.  Incidentally: Romina Abachi was an undergraduate Engineering student who had developed an interest in variable stars in high school.  Her physics teacher was an avid amateur astronomer and variable star observer!

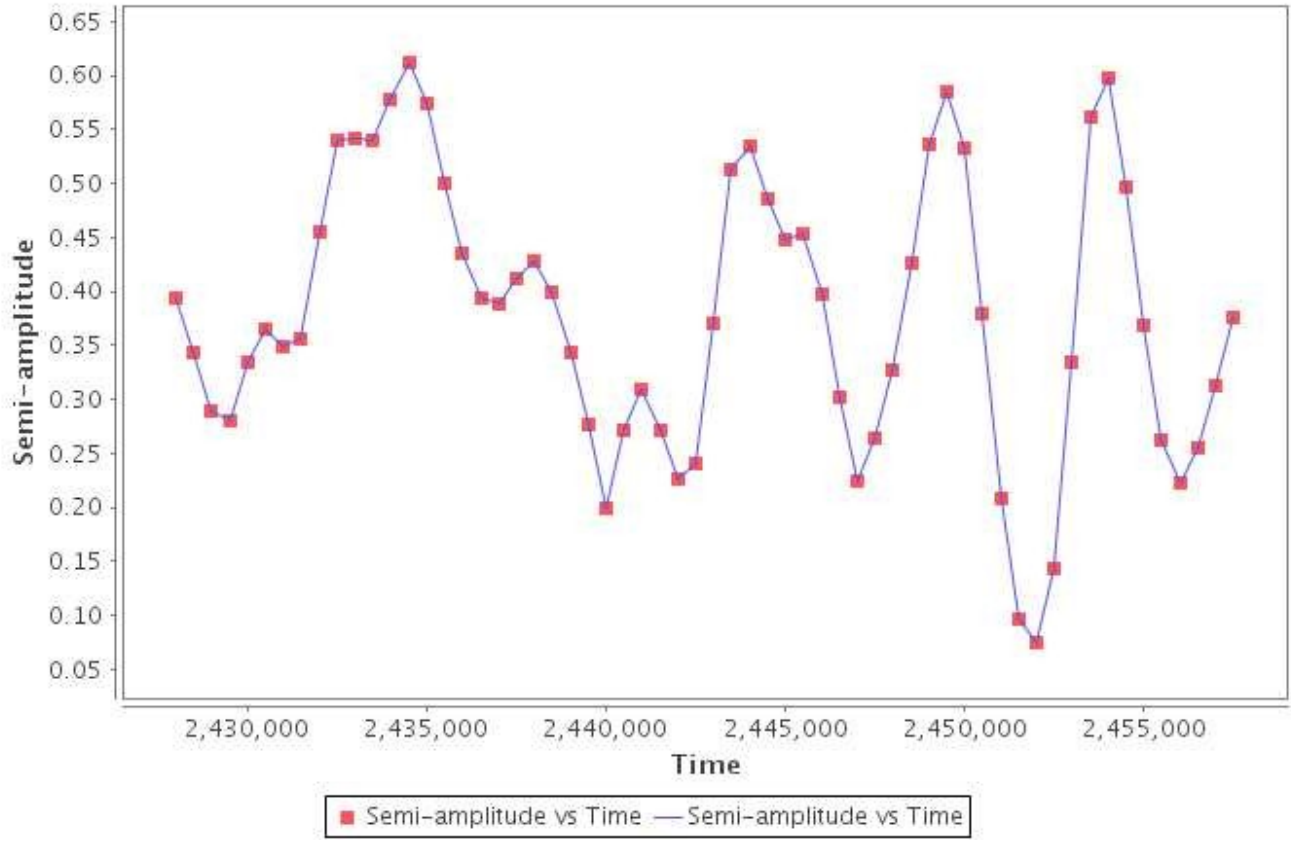

Figure 4: The variable pulsation amplitude of the pulsating red giant RV And, pulsation period 172 days (Percy & Abachi, 2013).

**Some Thoughts about the Future**

At the RTRSE conference, I was asked about the future of my program. Given that I have been "retired" for ten years, the future of my program is not infinite. But there is still science to be extracted from the AAVSO archive. This archive is being extended at both ends: new measurements are being made and added, and century-old Harvard photographic measurements are being digitized by the DASCH project, and added at the beginning of the archive. There are other other archives, such as PEP photometry obtained by robotic and survey telescopes: MACHO, OGLE, ASAS, and the AAVSONet epoch photometry. Volumes of data will come from new surveys such as the Large Synoptic Survey Telescope (LSST).

Many students are excited by the ideas of "big data", machine learning, and artificial intelligence. Interest in and experience with these will certainly lead to challenging, well-paying careers. Perhaps this is a way of drawing students from the mathematical sciences into astronomy.

"Big data" is also a possible area of interest for amateur astronomers. Many amateurs were at the forefront of the PC revolution, and the robotic telescope revolution. Many are professionals in the mathematical and computational sciences.

One drawback of my archival-research projects is that very few students actually make observations.

They do not get an "eyes-on" connection with the universe. Ideally, students should have the opportunity to make observations, and add these to the archive that they and other researchers will be using.

One of the many useful things that amateur astronomers and students can do is to comb through these archives, and identify variable stars whose behavior is unusual, and can therefore potentially tell us things that we do not know. The PRG research described above demonstrates that there are still unsolved problems in variable star astronomy, which can be found by mining the archives. The popularity of the *Galaxy Zoo* project indicates that many people are motivated by the possibility of classifying objects as normal or peculiar.

Could my program be scaled up? For a class, it would be easy to assign each student a carefully-chosen variable star to "solve". The class could then pool the results, as a team project, to draw overall conclusions. Or they could individually study groups of variable stars from the massive surveys mentioned above, and identify the ones which are worthy of further study.

*Acknowledgements*

I thank the organizers of the RTRSE conference for inviting me, and enabling me to participate "remotely"; the Natural Sciences and Engineering Research Council of Canada (NSERC) and the Ontario and University of Toronto Work-Study Programs for financial support; the organizers of my university's Research Opportunity Program, Mentorship Program, and Work-Study Program; the AAVSO observers and headquarters staff; and, most of all, the students who have worked with me over the last four decades.

**References**

AMARAL, A., & PERCY, J.R., 2016, An Undergraduate Research Experience on Studying Variable Stars. Journal of the American Association of Variable Star Observers (JAAVSO), 44, 72-77.

EDDINGTON, A.S., & PLAKIDIS, S., 1929, Irregularities of Period of Long-Period Variable Stars. Monthly Notices of the Royal Astronomical Society, 90, 65-71.

MATTEI, J.A., & PERCY, J.R., 2000, *Hands-On Astrophysics:* Variable Stars in Math, Science, and Computer Education. In: Amateur-Professional Partnerships in Astronomy, edited by J.R. Percy & J.B. Wilson, Astronomical Society of the Pacific Conference Series, Vol. 220, 296-298.

NEILSON, H.R., PERCY, J.R., & SMITH, H.A., 2016, Period Changes and Evolution in Pulsating Variable Stars. JAAVSO, 44, 179-195.

PERCY, J.R., DESJARDINS, A., & YEUNG, D., 1996, Be Stars in the AAVSO Photoelectric Program. JAAVSO, 25, 14-20.

PERCY, J.R. & COLIVAS, T., 1999, Long-Term Changes in Mira Stars. I. Period Fluctuations in Mira Stars. Publications of the Astronomical Society of the Pacific (PASP), 111, 94-97.

PERCY, J.R., WILSON, J.B., & HENRY, G.W., 2001, Long-Term VRI Photometry of Small-Amplitude Red Variables. I. Light Curves and Periods. PASP, 113, 983-996.

PERCY, J.R., HARLOW, C.D.W., & WU, A.P.S., 2004, Short-Period Variable Be Stars Discovered or Confirmed through Self-Correlation Analysis of *Hipparcos* Epoch Photometry.  PASP, 116, 178-183.

PERCY, J.R., 2006, Variable Star Research Projects for Outstanding Senior High School Students.  JAAVSO, 35, 284-287.

PERCY, J.R., 2007, Understanding Variable Stars (Cambridge UK: Cambridge University Press).

PERCY, J.R., 2008, Variable Star Research Experiences for High School Students, Undergraduates, and Amateur Astronomers.  In: Preparing for the 2009 International Year of Astronomy, edited by Michael G. Gibbs, Jonathan Barnes, James G. Manning, and Bruce Partridge, Astronomical Society of the Pacific  Conference Series 400, 363-367.

PERCY, J.R., ESTEVES, S., GLASHEEN, J., LIN, A., LONG, J., MASHINTSOVA, M., TERZIEV, E., & WU, S., 2010, Photometric Variability Properties of 21 T Tauri and Related Stars from AAVSO Visual Observations.  JAAVSO, 38, 151-160.

PERCY, J.R., & TERZIEV, E., 2011, Studies of "Irregularity" in Pulsating Red Giants. III. Many More Stars, an Overview, and Some Conclusions.  JAAVSO, 39, 1-9.

PERCY, J.R., 2012, The AAVSO Photoelectric Photometry Program in its Scientific and Socio-Historic Context.  JAAVSO, 40, 109-119.

PERCY, J.R., & ABACHI, R., 2013, Amplitude Variations in Pulsating Red Giants.  JAAVSO, 41, 193-213.

PERCY, J.R., & KHATU, V.C., 2014, Amplitude Variations in Pulsating Red Supergiants.  JAAVSO, 42, 1-12.

PERCY, J.R., & KIM, R.Y.H., 2014, Amplitude Variations in Pulsating Yellow Supergiants.  JAAVSO, 42, 267-279.

PERCY, J.R., 2015, Spurious One-Month and One-Year Periods in Visual Observations of Variable Stars.  JAAVSO, 43, 223-226.

PERCY, J.R., & DEIBERT, E., 2016, Studies of the Long Secondary Periods in Pulsating Red Giants.  JAAVSO, 44, 94-100.

WOOD, P.R., 2000, Variable Red Giants in the LMC: Pulsating Stars and Binaries? Publications of the Astronomical Society of Australia, 17, 18-21.